\documentclass[a4,twoside,12pt]{article}

\usepackage{color}
\usepackage{epsfig}

\begin{document}

\title{
Ivantsov parabolic solution for two combined moving interfaces}

\medskip

\author{D. Temkin \\[0.5cm]
Institut f\"ur Festk\"orperforschung, Forschungszentrum J\"ulich, \\
D-52425 J\"ulich, Germany \\
Contact e-mail: d.temkin@gmx.de}

\maketitle

\begin{abstract}
We demonstrate that for a migration of a
liquid layer between the melting and the solidification front
an exact steady-state solution with two parabolic fronts 
can be found. A necessary condition hereby is that the temperature
of the solidification front exceeds the temperature of the 
melting front (both temperatures are supposed to be constant). 
It is shown that in pure materials and alloys there exist
two types of solutions with two convex and with two concave
parabolas respectively. 
While a steady-state process with two planar interfaces is only
possible for a single point, the processes with two parabolas
are possible inside a region of control parameters. 
The relations between the Peclet numbers and the control
parameters are obtained.
\end{abstract}

\newpage

\section{Introduction} 
Liquid film migration (LFM) is well-known phenomenon 
which has been observed in many alloy systems during 
sintering in the presence of liquid phase \cite{yoon1,yoon2} 
and in Cu-In solid solutions during melting started at grain 
boundaries \cite{musch1}. In LFM, one of crystals is melted 
and the other one is solidified. The both solid-liquid interfaces 
are moving together with the same velocity. In the investigated 
alloys systems the migration velocity is of the order of $10^{-8}-10^{-7}$ 
$m\cdot s^{-1}$ and it is controlled  by the solute diffusion through 
a thin liquid layer between the two interfaces \cite{4}. The migration 
velocity is much smaller than the characteristic velocity of 
atomic kinetics at the interfaces. Therefore the both solids at the 
interfaces must be at the local thermodynamic equilibrium with 
liquid. The theory \cite{4} answers the question about 
the different equilibrium states at the melting and solidification fronts. 
In a steady-state regime, the difference in equilibrium relates to coherency 
stresses appearing only at the melting front due to the sharp profile of 
composition ahead the moving melting front. Thus, the liquid composition
at the melting front which depends on the coherency strains 
and a curvature of the front differs from the liquid composition 
at unstressed and curved solidification front. The migration velocity is 
proportional to the difference of these compositions divided by the 
film thickness \cite{4}. But what controls the thickness? 

Consequently, a new problem of two combined moving solid-liquid 
interfaces with a liquid between them appears. In the present article, 
this problem is considered under simplified boundary conditions: 
the temperature and chemical composition along each interface 
are constant. These constants are different for the melting and 
solidification fronts and differ from those far from the migrating 
liquid film. It means that any capillary, kinetic and crystallographic 
effects at the interfaces are neglected. It is found that under these simplified 
boundary conditions two co-focal parabolic fronts can move together 
with the same velocity. The situation is rather similar to a steady-state  motion 
of one parabolic solidification front into a supercooled melt \cite{1,2} 
or one parabolic melting front into a superheated solid. 

\section{Solutions for one parabolic front}
Needle-like stationary solutions were first obtained by Ivantsov 
for the crystallization of a pure material from a supercooled
melt \cite{1} and were extended to binary alloys \cite{2}. 
This solutions describes a parabolic interface of a solid phase
at constant temperature $T_m$ which extends into a 
supercooled melt. Inside the melt the temperature drops 
and reaches far from the interface its asymptotic value $T_{0} < T_m$. 

For the two-dimensional case the Ivantsov relation is
\begin{equation}
  \Delta = F(P) \equiv \sqrt{\pi P} e^P {\rm erfc} (\sqrt{P}) 
  \quad , \quad 0 \le \Delta < 1.
  \label{ivantsov1}
\end{equation}
It connects the Peclet number $P = VR/2D$ and the supercooling
$\Delta = (T_m-T_{0}) c/q$. Here $V$ and $R$ are the front-velocity
and the tip-radius of the parabola respectively; $D$ and $c$ are
the thermal diffusivity and the specific heat of the melt, respectively, $q$ is
the latent heat. 

Eq. (\ref{ivantsov1}) can easily be obtained in 2D-parabolic
coordinates $(\eta,\xi)$ (see, for example, Ref. \cite{3}) 
\begin{eqnarray}
  \eta &=& \sqrt{x^2+z'^2} + z' \quad , \quad \xi =
  \sqrt{x^2+z'^2} - z' \quad, \quad ß0 \leq (\eta, \xi) 
  < \infty, \nonumber \\ x &=& \pm \sqrt{\eta \xi} \quad , 
  \quad z' = \frac{1}{2} \ (\eta - \xi) \quad , 
  \quad - \infty < (x,z') < \infty,
\end{eqnarray}
where $(x,z'=z-Vt,t)$ are the Cartesian coordinates and the $t$ is the time.
In this description the parabolic solidification front
($z'\!=\!\frac{R}{2}\,-\,\frac{x^{2}}{2R}$), 
moving with velocity $V$ in the positive $z$-direction,
has the coordinate $\eta = R$. The regions $0 \leq \eta < R$ and 
$R < \eta < \infty$ correspond to the solid and the liquid phase
respectively.

In this set of coordinates, the temperature field $T$ in both phases
depends only on $\eta$. The thermal diffusion equation 
\begin{equation}
  \frac{\partial}{\partial \eta} \ \left(  \sqrt{\eta} \ \frac{\partial T}
  {\partial \eta}\right) + \frac{V}{2D} \left( \sqrt{\eta} \ 
  \frac{\partial T}{\partial \eta} \right) \ = \ 0
  \label{diffusion1}
\end{equation}
is easily solved by the ansatz
\begin{equation}
T (\eta) = A+B \int \eta^{-1/2} e^{-V\eta/2D} d\eta
\end{equation}
with different constants $A$ and $B$ for the solid and the
liquid phase. These constants and Eq. (\ref{ivantsov1}) can
be obtained by using the appropriate boundary conditions,
namely the asymptotic boundary condition
$T (\eta \to \infty) = T_{0}$ and the interfacial conditions
of temperature continuity $T(R-0)=T(R+0)=T_m$, and 
heat balance at the interface
\begin{equation}
  V q/c = - 2D T' (R+0) + 2DT'(R-0),
\end{equation}
where the prime denotes the derivative with respect to 
$\eta$. 
For reasons of simplicity we assumed the same values for
$D$ and $c$ in both phases.

Apart from the solution with convex parabolic front there
exists as well a solution with a concave front 
($z'\!=\!-\frac{R}{2}+\frac{x^2}{2R}$).
In the parabolic coordinates, the interface is defined by the  
relation $\xi = R$ while solid and liquid phase are at 
$\infty < \xi < R$ and  $R > \xi \ge 0 $, respectively.
In contrast to the former case the temperature 
field $T$ depends only on $\xi$. Instead of 
Eq. (\ref{diffusion1}) the diffusion equation therefore
reduces to
\begin{equation}
  \frac{\partial}{\partial\xi} \ \left( \sqrt{\xi} \ 
  \frac{\partial T}{\partial \xi} \right) \ - \
  \frac{V}{2D} \left( \sqrt{\xi} \ \frac{\partial T}
  {\partial \xi} \right) \ = \ 0 \qquad ,
  \label{diffusion2} 
\end{equation}
and is solved by the ansatz
\begin{equation}
  T(\xi) \ = \ A + B \, \int \xi^{-1/2} e^{V\xi/2D} d\xi \qquad.
\end{equation}
While the conditions of continuity and heat balance at the interface
stay the same as for the convex parabola,  
the asymptotic condition changes to $T (\xi \to 0) = T_{0}$. 
Thus for the concave front instead of Eq. (\ref{ivantsov1})
we obtain the relation
\begin{equation}
  \Delta = \Phi (P) \equiv 2 \sqrt P \ e^{-P} \ 
  \int^{\sqrt P}_0 e^{x^2} dx \qquad \quad 0 \leq \Delta \leq 1.284... \qquad , 
\label{ivantsov2} 
\end{equation}
between the supercooling $\Delta$ and the Peclet number $P$.
The function $\Phi (P)$, in contrast to the function $F (P)$ 
in Eq. (\ref{ivantsov1}), is not monotonous: it has a maximum 
$1.284...$ at $P = 2.25...$ and
approaches 1 from above for $P\to\infty$.

Naturally, Eq. (\ref{ivantsov1}) and Eq. (\ref{ivantsov2}) describe 
as well a parabolic melting front: the superheated solid
phase with $T_{0} > T_m$ is outside the convex parabola 
($\eta = R$) at $R<\eta<\infty$ or inside the concave one
($\xi=R$) at $R>\xi\geq 0$. The melting heat is $-q$. 

\section{Two parabolic fronts in a pure material}
It turns out, that an exact steady-state solution with a combined
motion of the two co-focal parabolic fronts 
(melting and solidification front with a liquid layer in between)
can be found. For this the solidification front temperature 
$T_m$ has to exceed the temperature $\tilde T_m$ on the melting front. 
This may be the case for pure materials where the melted solid
phase (S1) has with respect to the solidified solid (S2) an 
additional contribution $\epsilon$ to the energy-density 
(for example due to preliminary mechanical treatment). 
In this case the melting heat equal to
$- (q - \epsilon)$ and the melting temperature 
$\tilde T_m = T_m (1-\epsilon/q)$  differ from the values 
$q$ and $T_m$ for the solidification front. 
The additional energy $\epsilon$ is the driving force for a
recrystallization process, which takes place in the 
solid either due to a migration of a boundary between 
a S1- and a S2-grain or due to a possible recrystallization 
through a liquid layer between S1 and S2. 

It should be pointed out, that a combined stationary motion
of two planar interfaces (with equilibrium temperatures $T_m$ and
$\tilde T_m$) is only possible at the temperature $T_{0,0}$
at which the temperature increase $(T_m - T_{0,0})$ exactly
compensates the additional energy, $c(T_m - T_{0,0})= \epsilon$. 
All temperatures deviating from $T_{0,0}$ lead to an increase 
(at $T_{0} > T_{0,0}$) or to a decrease and disappearance 
(at $T_{0} < T_{0,0}$) of the liquid layer between the planar
interfaces. Here only two co-focal parabolic interfaces can lead to a
steady-state motion. 

Similar to the considerations described for one interface
an analysis of the temperature field in parabolic coordinates
leads to the following relations:\\
For two convex parabolas with $P_1 > P_2$ (Fig. 1):
\begin{eqnarray}
  \Delta_m &=& 2 \sqrt{P_2} e^{P_2}
  \ \int^{\sqrt{P_1}}_{\sqrt{P_2}} e^{-x^2} \ dx \label{ivantsovneu1a}\\
  \Delta~~ &=& \left[ 1-\alpha \Delta_m - \sqrt{\frac{P_2}{P_1}} \ 
  e^{P_2 - P_1} \right] \ F(P_1). \label{ivantsovneu1b}
\end{eqnarray}
For two concave parabolas with $P_1 < P_2$ (Fig. 1):
\begin{eqnarray}
  \Delta_m &=& 2 \sqrt{P_2} e^{-P_2} \ 
  \int^{\sqrt{P_2}}_{\sqrt{P_1}} e^{x^2} \ dx, \label{ivantsovneu2a}\\
  \Delta~~ &=& \left[ 1-\alpha \Delta_m - \sqrt{\frac{P_2}{P_1}} \ 
    e^{P_1 - P_2} \right] \ \Phi (P_1), \label{ivantsovneu2b}
\end{eqnarray}
with $P_i = V R_i/2D$, $\Delta_m = (T_m - \tilde T_m) c/q$,
$\Delta = (T_{0} - \tilde T_m) c/q$ and $F(P)$ and $\Phi (P)$
as in Eq. (\ref{ivantsov1}) and Eq.(\ref{ivantsov2}). 
$\alpha = q/c T_m$ is a material parameter (e.g., for Ni one has 
$\alpha \cong 0,25$).
The term $(1-\alpha \Delta_m)$ arises due to the fact that the 
melting heat of a solid phase S1 with an additional 
energy density $\epsilon$ differs from the equilibrium 
melting heat, $-q$ 
(note that $-(q-\epsilon) = -q\,(1-\alpha \Delta_m)$). 

In the corresponding limiting cases Eqs. 
(\ref{ivantsovneu1a})-(\ref{ivantsovneu2b}) reduce to 
Eqs. (\ref{ivantsov1}) and (\ref{ivantsov2}):
Eq. (\ref{ivantsovneu1a}) for $P_1\to\infty$ and  Eq. (\ref{ivantsovneu2a}) 
for  $P_1 \to 0$
 lead to the solidification relations  Eq. (\ref{ivantsov1}) and
Eq. (\ref{ivantsov2}), respectively, with supercooling $\Delta_m$ instead of $\Delta$.
Eq. (\ref{ivantsovneu1b}) for $P_2\to 0$ and 
Eq. (\ref{ivantsovneu2b}) for $P_2\to\infty$
lead to the melting relations  Eq. (\ref{ivantsov1}) and
Eq. (\ref{ivantsov2})
with the normalized superheating 
 $\Delta/(1-\alpha \Delta_m)$ instead of $\Delta$.

\section{Analysis of convex and concave solutions}
Even if Eqs. (\ref{ivantsovneu1a}) - (\ref{ivantsovneu2b}) 
are valid for arbitrary values of the normalized 
driving force $\Delta_m > 0$, we consider in the
following analysis the case of small
driving forces $\Delta_m \ll 1$. With this assumption 
Eqs. (\ref{ivantsovneu1a}) and (\ref{ivantsovneu2a}), respectively, simplify to 
\begin{eqnarray}
  \sqrt{P_2} &\cong& \frac{1}{2} \ \left[ \sqrt{P_1} \pm 
  \sqrt{P_1 - 2\Delta_m} \right] \quad \quad (P_1 > P_2)~
  \label{ivantsov3}\\
  \sqrt{P_2} &\cong& \frac{1}{2} \ \left[ \sqrt{P_1} + 
  \sqrt{P_1 + 2\Delta_m} \right] \quad  \quad (P_1 < P_2).
  \label{ivantsov4}
\end{eqnarray}
Eq. (\ref{ivantsov3}) together with Eq. (\ref{ivantsovneu1b}) and 
Eq. (\ref{ivantsov4}) together with Eq. (\ref{ivantsovneu2b})
define for the case of small driving forces $\Delta_m \ll 1$
the dependency of $\Delta (P_1)$ for both types of solutions 
(concave and convex parabolas) (see Fig. 2).

For  $P_1 \gg \Delta_m$ 
 the upper branch of Eq. (\ref{ivantsov3}) gives  
\begin{equation}
  P_2 \cong P_1 - \Delta_m \qquad   (P_1 > P_2)
\end{equation}
and from Eq. (\ref{ivantsov4})
\begin{equation}
  P_2 \cong P_1 + \Delta_m \qquad (P_1 < P_2).  
\end{equation}
Using this simplification and taking the limit $\Delta_m \to 0$ we 
obtain from Eqs. (\ref{ivantsovneu1b}) and  (\ref{ivantsovneu2b})
the simplified expressions
\begin{eqnarray}
  \frac{\Delta_{~~}}{\Delta_m} &=& \left( 1-\alpha \, + \, 
  \frac{1}{2P_1} \right) \ F(P_1) \quad , \quad 
  (P_1 > P_2) \qquad , \label{exequation20} \\
  \frac{\Delta_{~~}}{\Delta_m} &=& \left( 1-\alpha \, - \, 
  \frac{1}{2P_1} \right) \ \Phi (P_1) \quad , \quad
  (P_1 < P_2) \qquad . \label{exequation21}
\end{eqnarray}

A stationary solution with two convex parabolas $(P_1 > P_2)$ exists for
\begin{equation}
  \Delta_{0} \leq \Delta \leq \Delta^*  \label{regiondef1}
\end{equation}
with 
\begin{equation}
  \Delta_{0} \equiv (1-\alpha) \, \Delta_m \qquad\mbox{and}\qquad
  \Delta^* \equiv 1 - \alpha \Delta_m  .
\end{equation}
The point $\Delta = \Delta_{0}$ corresponds to the 
steady-state solution of two moving planar interfaces mentioned above.

If $\Delta_{0} \leq \Delta \ll \Delta_1$, the dependence of 
$\Delta$ on $P_1$ is defined by Eq. (\ref{exequation20}) 
and $P_2 \cong P_1 - \Delta_m$. At the point 
$\Delta = \Delta_1 \cong \sqrt{\pi \Delta_{m}/2}$ 
one has $P_1 \cong 2 \Delta_m$ and $P_2 \cong \Delta_m/2$. 
In the vicinity of the $\Delta_1$-point one can obtain 
\begin{equation}
  \frac{\Delta-\Delta_1}{\Delta_1} \ \cong \ \mp
  \sqrt{\frac{P_1}{2\Delta_m} - 1}. 
\end{equation}

From Eqs. (\ref{ivantsovneu1a}) and (\ref{ivantsovneu1b}), one can 
obtain that close to the limiting point $\Delta^*$ one gets 
\begin{equation}
  P_1 \ \cong \ \frac{\Delta^*}{2(\Delta^*-\Delta)} \ \gg \ 1 \quad 
  and \quad P_2 \ \cong \ \frac{\Delta_{m}^2}{\pi} \ll 1.
\end{equation}
These relations describe the independent motion of both parabolic 
fronts propagating with the same velocity.

Solutions of Eqs. (\ref{ivantsovneu2b}) - (\ref{ivantsov4})
 for a pair of concave parabolas exist in the region 
\begin{equation}
  -\mid\Delta_4 \mid\ \leq\ \Delta\ \leq\ \Delta_2 \label{regiondef2}
\end{equation}
between two extreme points on the $\Delta (P_1)$-dependence
(Fig. 2): minimum point $\mid \Delta_4 \mid \ \cong \ \Delta_m$ 
at $P_1 \cong \sqrt{\Delta_m/4 (1-\alpha)}$ and maximum point
 $\Delta_2$ which is close to $\Delta_{0}$. In the limit $P_1 \to
 \infty$ the value of $\Delta$ tends to the same point $\Delta_{0}$
 as for the convex parabolas, $\Delta \cong \Delta_{0} -\alpha/2P_1$.

\section{Parabolic solutions for binary alloys}
Extending our analysis to a two-component alloy, we have to take into
account that in contrast to a pure material the temperatures 
$T_1$ and $T_2$ of the melting and solidification fronts are 
unknown (in a pure material $T_1 = \tilde T_m$, $T_2 = T_m$). 
In order to define $T_1$ and $T_2$ (and  consequently 
$\Delta_m = (T_2 - T_1) c/q$, and $\Delta = (T_{0} - T_1) c/q$ 
in Eqs. (\ref{ivantsovneu1a})-(\ref{ivantsovneu2b})) we have
to obtain two additional equations.

In our description, we denote the molar fraction of the
second component with $C$ and the diffusion 
coefficients in the solid and the liquid phase with 
$D_S$ and $D_L$ respectively. While for convex parabolas
the concentration fields, $C(\eta,\xi)$, only depend on $\eta$,
they depend only on $\xi$ in the concave case. 
In the first case, the field $C(\eta)$ satisfies the
equilibrium boundary conditions at both interfaces, 
namely the continuity condition
\begin{equation}
  C(R_{2}-0) = C_S (T_2) \ , \ 
  C(R_{2}+0) = C_L (T_2) \ , \
  C(R_{1}-0) = \tilde C_L (T_1) \ , \
  C(R_{1}+0) = \tilde C_S (T_1) \quad , \label{alloycond1}
\end{equation}
the far field condition $C(\eta\!\to\!\infty) = C_{0}$
and the conservation conditions at the interfaces
\begin{eqnarray}
  V \left[ C_L(T_2) - C_S(T_2)\right] &=& -2D_L C' (R_2+0) +
  2D_S C' (R_2-0) \quad ,  \label{alloycond2}\\
  V \left[ \tilde C_L(T_1) - \tilde C_S(T_1)\right] &=&
  -2D_L C' (R_1-0) + 2D_S C' (R_1+0) \quad . \label{alloycond3}
\end{eqnarray}

Here $C_S (T_2)$ and $C_L (T_2)$ are the solidus and liquidus
compositions which are defined by the equilibrium phase 
diagram of the alloy, while $\tilde C_S(T_1)$,
$\tilde C_L (T_1)$ are the corresponding values defined 
by a disturbed phase diagram. The additional
energy density $\epsilon$, changing the melting point of the melted
solid phase S1, will change the equilibrium compositions at 
the melting interface as well. In this case, the composition 
differences $\tilde C_S (T_1) - C_S (T_1)$, $\tilde C_L (T_1)
- C_L (T_1)$, are proportional to $\epsilon/q$. Another reason
for the distortion of the phase diagram are coherency stresses
due to compositional inhomogeneities in front of the melting 
interface \cite{4}.

By solving the diffusion equation for $C(\eta)$ and applying the
appropriate boundary conditions (\ref{alloycond1})-(\ref{alloycond3}),
we obtain for the case of convex parabolas the solution:
\begin{eqnarray}
  \frac{C_L (T_2) - \tilde C_L (T_1)}{C_L (T_2) - C_S (T_2)} \!&\!=\!&\!
  2\sqrt{P_{2L}} e^{P_{2L}} \ \int^{\sqrt{P_{1L}}}_{\sqrt{P_{2L}}}
  e^{-x^2} dx \quad , \quad P_{1L} > P_{2L} \label{alloysol1a}\\
  \frac{C_{0}-\tilde C_{s}(T_1)}{C_L(T_2)-C_{s}(T_{2})}\!&\!=\!&\!
  \left[
     \frac{\tilde C_L (T_1) - \tilde C_s (T_1)}{C_L (T_2) - C_s (T_2)}
     -\sqrt{\frac{P_{2L}}{P_{1L}}}\,e^{P_{2L} - P_{1L}} 
  \right] \ F(P_{1S}) \label{alloysol1b}
\end{eqnarray}
($P_{iL} = DP_i / D_L$, $P_{1S} = D_L P_{1L} / D_S$; $F(P)$ 
is the same function as in Eq. (\ref{ivantsov1})). A similar
consideration for the case of concave parabolas with 
concentration field $C(\xi)$ leads to:
\begin{eqnarray}
  \frac{ C_L (T_2) - \tilde C_L (T_1)}{C_L (T_2) - C_s (T_2)} \!&\!=\!&\!
  2 \sqrt{P_{2L}} \ e^{-P_{2L}} \ \int^{\sqrt{P_{2L}}}_{\sqrt{P_{1L}}} 
  e^{x^2} dx \quad , \quad P_{1L} < P_{2L} \label{alloysol2a}\\
  \frac{C_{0}-\tilde C_{s}(T_1)}{C_L(T_2)-C_{s}(T_{2})}\!&\!=\!&\!
  \left[ \frac{\tilde C_L (T_1) - \tilde C_s (T_1)}{C_L (T_2) - C_s (T_2)}
     -\sqrt{\frac{P_{2L}}{P_{1L}}}\,e^{P_{1L} - P_{2L}} 
  \right] \ \Phi(P_{1S}), \label{alloysol2b}
\end{eqnarray}
where $\Phi(P)$ is defined by the same manner as 
in Eq. (\ref{ivantsov2}). 


For the convex configuration, Eqs. 
(\ref{ivantsovneu1a})-(\ref{ivantsovneu1b}) together with Eqs. 
(\ref{alloysol1a})-(\ref{alloysol1b}) describe the dependency
of the quantities $T_1$, $T_2$, $P_1$, $P_2$ on the initial
control parameters $T_{0}$, $C_{0}$, and $\epsilon$.
The corresponding relations for the concave configuration are given by 
Eqs. (\ref{ivantsovneu2a})-(\ref{ivantsovneu2b}) and 
Eqs. (\ref{alloysol2a})-(\ref{alloysol2b}).
In systems with vanishing $\epsilon$ the coherency strain effect
can support the combined motion of two interfaces as, for example,
in liquid film migration \cite{4}.

\section{Discussion} 

The considered two fronts process (TFP) is an alternative to an one front process (OFP).
TFP is possible at some conditions (for example, at $\tilde T_m < T_m$ for a pure 
material) and do exist as a steady-state process in a definite region 
of control parameters.
 In this region there is one or a few OFP. For  pure materials the TFP exists at 
$-|\Delta_4|\le \Delta\le \Delta^{\ast}$ (see Eqs. (19) and (23)), i.e. at 
$\tilde T_m-(T_m-\tilde T_m)\le T_0\le \tilde T_m+(q-\epsilon)/c$.
When the initial temperature is $T_0<\tilde T_m<T_m$, the OFP of the transition 
$S1\rightarrow S2$ proceeds at the grain boundary $S1/S2$. 
When $\tilde T_m<T_0<T_m$, the OFP of the melting of $S1$ at the interface $S1/L$ 
is also possible. At $T_0>T_m$ the grain $S2$ can also be melted at the interface 
$S2/L$. The TFP can proceed faster than the corresponding OFP due to a heat transfer 
between melting and solidification fronts through a thin liquid layer. 
In order to initiate the TFP (especially at 
$ \tilde T_m-(T_m-\tilde T_m)\le T_0\le \tilde T_m$) a liquid phase must be created 
inside the system. Then the grain boundary $S1/S2$ splits into two interfaces, 
$S1/L$ and $S2/L$, and the TFP proceeds as a self-sustained process.

It should be noted that the TFP can take place in a pure material in which 
there are several polymorphic modifications. Then a low temperature modification plays 
a role of the melted grain $S1$ in Fig. 1 and its melting temperature $\tilde T_m$ 
is lower than the melting temperature $T_m$ for a high temperature modification, $S2$.
In such a case the solutions given by Eqs. (\ref{ivantsovneu1a}) - 
(\ref{ivantsovneu2b}) are valid for $\Delta_m=(T_m-\tilde T_m)c/q$ as a material 
parameter and $(1-\alpha\Delta_m)$  is replaced by another material parameter 
$\tilde q/q$, where $\tilde q$ and $q$ are the melting heats of the low and high 
temperature phases respectively. 

Generally, one can consider the TFP of the transition of a phase $S1$ into another 
phase $S2$ through an intermediate phase $L$ (which is not necessarily a liquid one) 
as an alternative to the OFP of the direct transition $S1\rightarrow S2$.      

The obtained Eqs. (\ref{ivantsovneu1a}) - (\ref{ivantsovneu2b}) 
together with Eqs. (\ref{alloysol1a}) - (\ref{alloysol2b}), 
which define relations between the Peclet numbers $P_1$, $P_2$ 
and control parameters $T_0$, $C_0$, give the continuous spectrum 
of solutions with the only free parameter (e.g., the velocity $V$). 
Therefore an additional equations, i.e. ``selection'' relationship, 
is needed in order to define the unique solution. This is similar 
to a well-known ``selection'' problem in dendritic growth \cite{5}. 
A search for the selection condition will be a subject of future 
investigation. The only point which might be stressed here 
is the following. 

The structure of the fronts is usually more complicated compared to pure parabolic 
ones due to possible cellular structures or due to finite size of the sample. 
In this case,
two parabolas describe only a part of the moving fronts. 
The complete shape of the fronts must be subjected to 
additional boundary conditions.  One can speculate on  
 possible cellular structures with two moving 
interfaces which are shown in Fig. 3. These structures should appear due to the 
diffusional interaction between different cells and capillary effects play also 
crucial role. 
The structures have 
central parabolic parts which are convex (Fig. 3(a,b)) 
or concave (Fig. 3(c)) and satisfy boundary conditions 
of zero heat- and mass-fluxes across boundaries of cells. 
The first structure (Fig. 3(a)) can be related to a melting processes
and the other two ones, for example, to sintering of two 
solid grains $S1$ and $S2$ in liquid phase $L$. 
The structure in Fig. 3(b) may correspond to a sintering process 
with a supersaturated solid $S1$, while
the structure in Fig. 3(c) with a concave central part 
may correspond to the case of undersaturated solid $S1$.

In the process in the channel (or cell) the fronts velocity may be controlled 
either by the above mentioned cell boundary conditions or by the ``selection''
which takes place mostly in the parabolic region. Such a problem 
arises also in the selection of the growth velocity of the 
classical dendrite in the channel (the structure with one front) \cite{5}. 

In this paper the solutions are obtained for two parabolic fronts 
in the two dimensions. Similar solutions can be easily found 
for two paraboloidal interfaces in the three dimensions. 
 
A few words about the solutions (\ref{ivantsov1}) and (\ref{ivantsov2}) 
for one convex and one concave parabolic interface might be outlined. 
The convex parabola and Eq. (\ref{ivantsov1}) are playing an important role 
in dendritic solidification (see, e.g., Ref. \cite{5}) and must be important 
in ``dendritic'' melting. It can be supposed that concave parabola 
and Eq. (\ref{ivantsov2}) are playing a role in such ``doublon'' 
structures with two convex parts and a concave part in between 
which are similar to the profile of $S2/L$-interface in Fig. 3(c). 
Whether or not those ``doublon'' structures exist in solidification 
or melting processes with one front? 

\section{Conclusions} 
A theoretical model for combined motion of two solid-liquid fronts with 
a liquid layer in between has been developed for pure materials 
and binary alloys. The temperature and chemical compositions 
at the interfaces were supposed to be constants but different for 
fronts of melting and solidification. 

An exact solution which describes the steady-state motion of two 
co-focal parabolic interfaces has been found. 
It is shown that there exist two types of solutions 
with two convex and with two concave parabolas. The relations 
between the Peclet numbers and the control parameters of the process 
for both types of solutions are obtained. 

As usual in theories of the steady-state growth, 
the continuous spectrum of solutions with one 
free parameter exists. 
The unresolved problem of ``selection'' of unique solution 
is briefly discussed. 

\bigskip

\section{Acknowledgments}
This work was supported in part by the Deutsche Forschungsgemeinschaft under project 
SPP 1120.
The author would like to thank L. Bagrova, E. Brener, P. Galenko, 
 and H. M\"uller-Krumbhaar for fruitful discussions.

\newpage
FIGURE CAPTIONS

Figure 1: Two types of combined moving parabolic fronts: convex (left) and 
concave (right).

Figure 2: The dependence of the reduced temperature $\Delta$ on the Peclet number
$P_1$ for convex and concave fronts. The insert represents the region of small
$\Delta$. Parameters are: $\Delta_m=0.01$ and $\alpha=0.25$.

Figure 3: Three possible cellular structures 
of two combined moving interfaces with convex [(a) and (b)] 
and concave (c) central parabolic part. $S1$ and $S2$ are 
melting solid and growing solid, and $L$ is the liquid phase. 
The direction of motion is defined by the arrow of $V$. 
Dashed regions in (a)-(c) show the central parabolic part of interfaces.

\newpage

\begin{figure}
\begin{center}
\epsfig{file=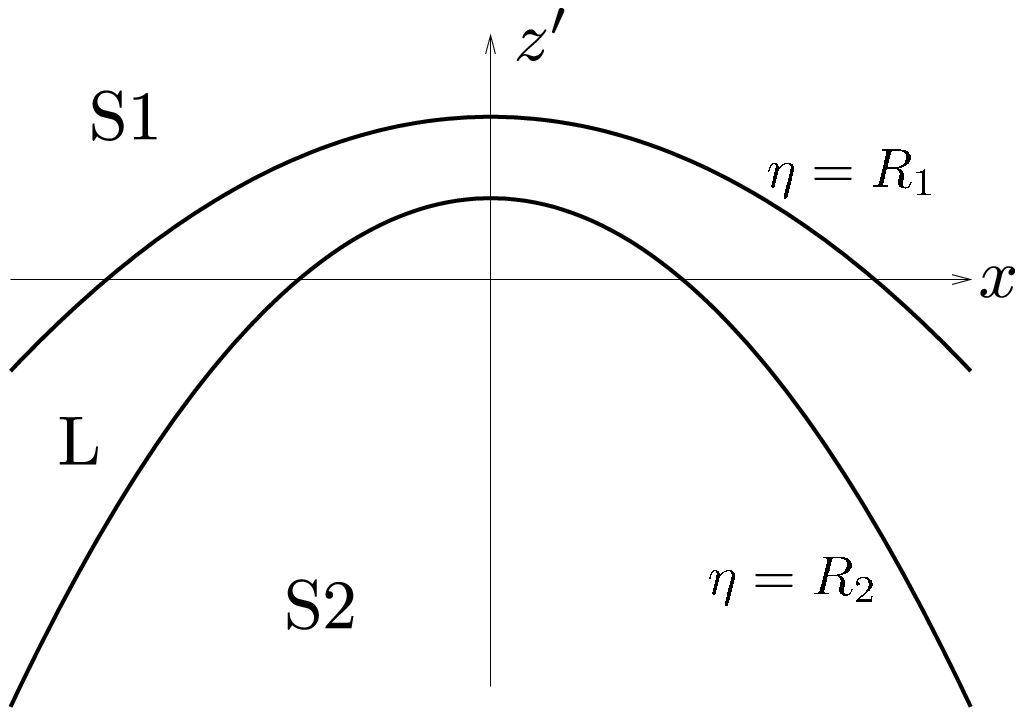, width=8cm}
\epsfig{file=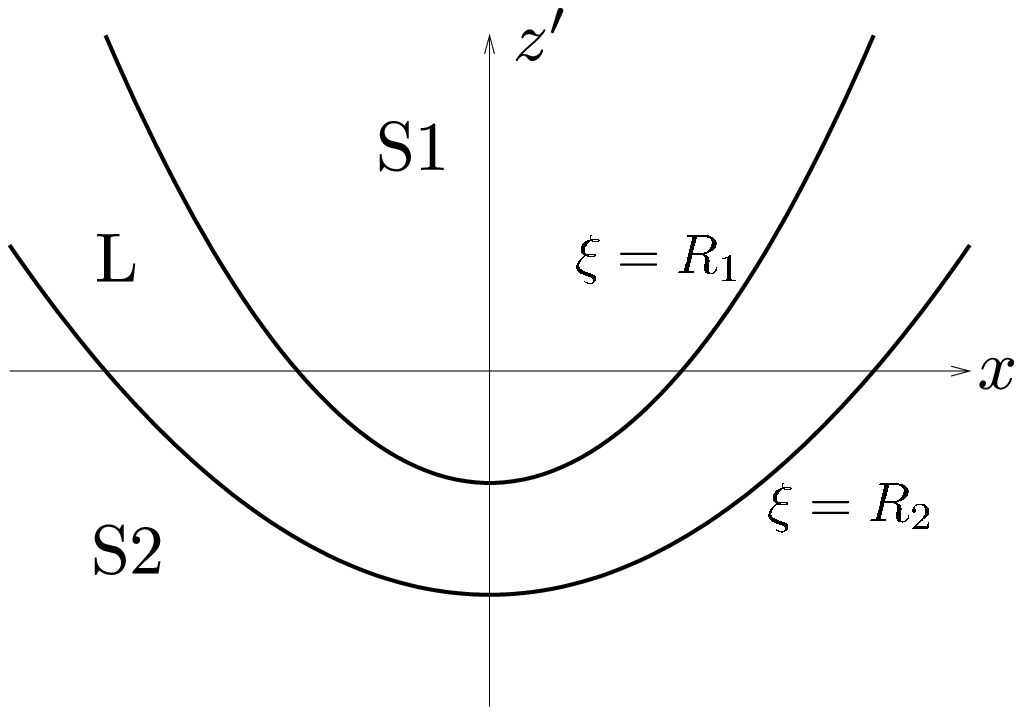, width=8cm}
\caption{	}
\end{center}
\end{figure}

\newpage

\begin{figure}
\begin{center}
\epsfig{file=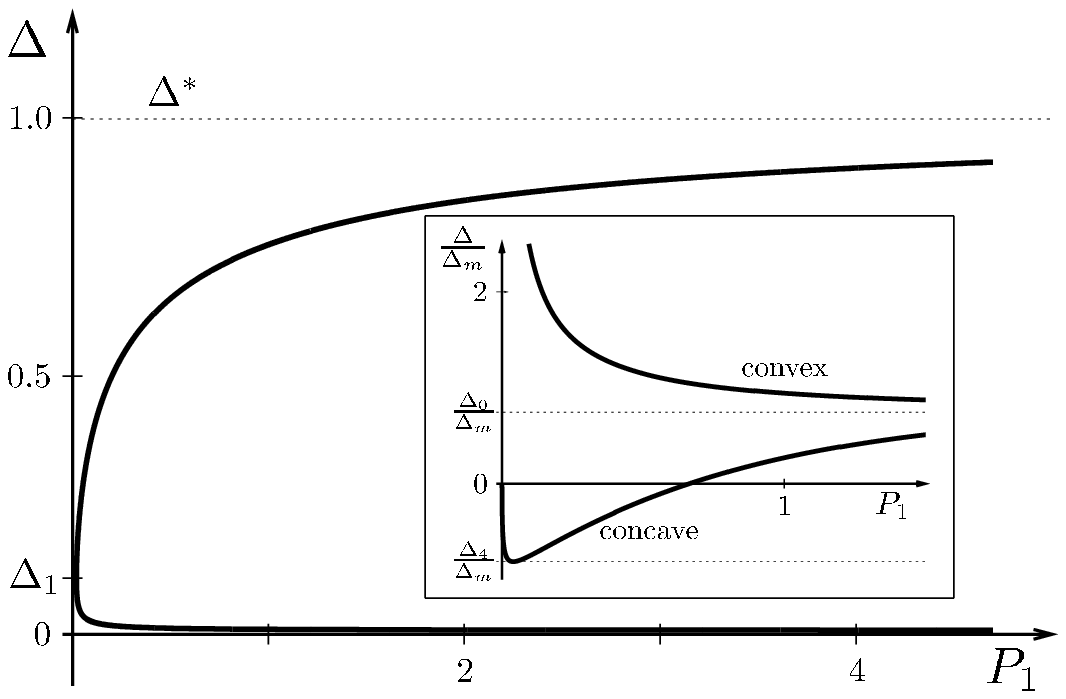, width=10cm}
\caption{	}
\end{center}
\end{figure}

\newpage

\begin{figure}
\begin{center}
\epsfig{file=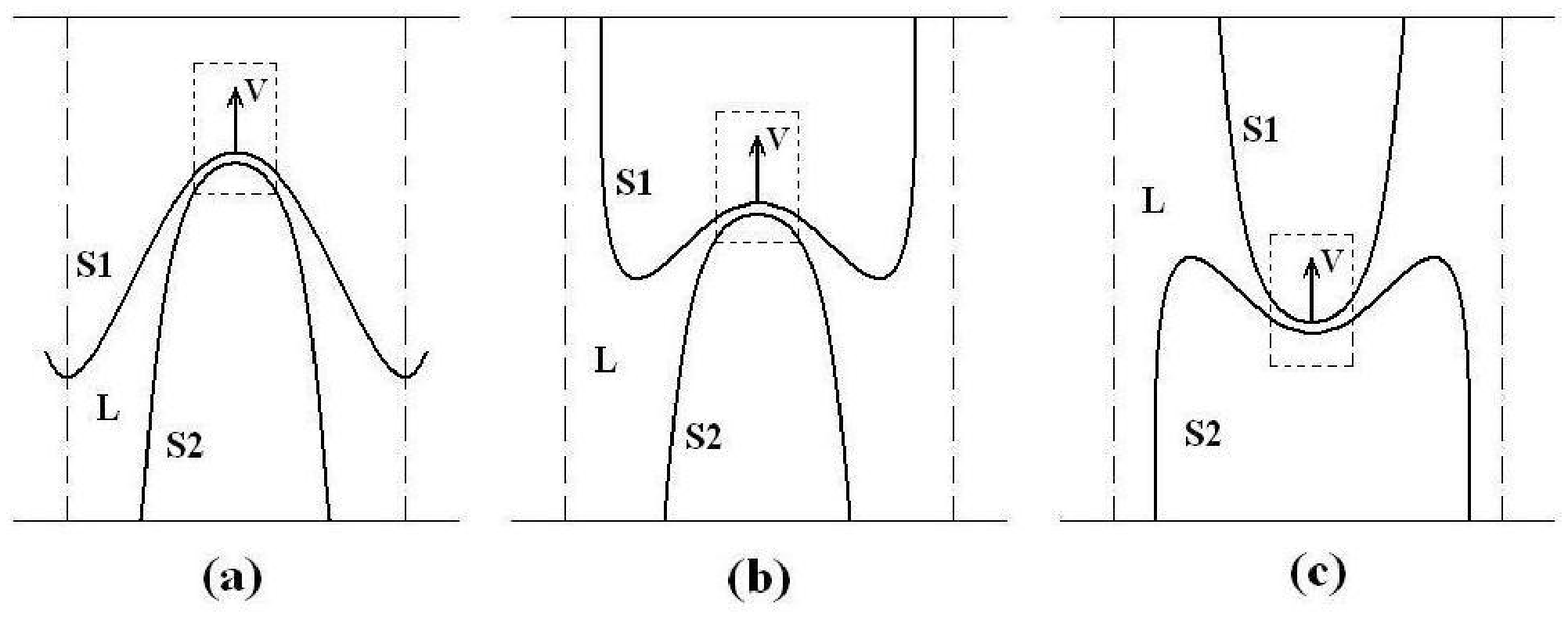, width=17cm}
\caption{	}
\end{center}
\end{figure}

\end{document}